\documentclass[aps,preprint,floats,epsf,epsfig,nofootinbib]{revtex4}
\usepackage{epsfig}

\begin{document}
\newcommand{\sla}[1]{#1\!\!\!\slash}
\def\be{\begin{eqnarray}}
\def\en{\end{eqnarray}}
\def\non{\nonumber}
\def\la{\langle}
\def\ra{\rangle}
\def\nc{N_c^{\rm eff}}
\def\vp{\varepsilon}
\def\drho{\bar\rho}
\def\deta{\bar\eta}
\def\CP{{\it CP}~}
\def\a{{\cal A}}
\def\B{{\cal B}}
\def\c{{\cal C}}
\def\d{{\cal D}}
\def\e{{\cal E}}
\def\p{{\cal P}}
\def\t{{\cal T}}
\def\up{\uparrow}
\def\dw{\downarrow}
\def\vma{{_{V-A}}}
\def\vpa{{_{V+A}}}
\def\smp{{_{S-P}}}
\def\spp{{_{S+P}}}
\def\J{{J/\psi}}
\def\ov{\overline}
\def\Lqcd{{\Lambda_{\rm QCD}}}
\def\pr{{Phys. Rev.}~}
\def\prl{{Phys. Rev. Lett.}~}
\def\pl{{Phys. Lett.}~}
\def\np{{Nucl. Phys.}~}
\def\zp{{Z. Phys.}~}
\def\lsim{ {\ \lower-1.2pt\vbox{\hbox{\rlap{$<$}\lower5pt\vbox{\hbox{$\sim$}
}}}\ } }
\def\gsim{ {\ \lower-1.2pt\vbox{\hbox{\rlap{$>$}\lower5pt\vbox{\hbox{$\sim$}
}}}\ } }


\font\el=cmbx10 scaled \magstep2{\obeylines\hfill October, 2006}

\vskip 1.5 cm \centerline{\large\bf Scalar Glueball, Scalar
Quarkonia, and their Mixing}
\bigskip
\centerline{\bf Hai-Yang Cheng$^1$, Chun-Khiang Chua$^{1\dagger}$
and Keh-Fei Liu$^{1,2}$}
\medskip
\centerline{$^1$ Institute of Physics, Academia Sinica}
\centerline{Taipei, Taiwan 115, Republic of China}
\medskip
\medskip
\centerline{$^2$ Department of Physics and Astronomy, University
of Kentucky} \centerline{Lexington, KY 40506}
\medskip
\bigskip
\bigskip
\centerline{\bf Abstract}
\bigskip
{\small

The isosinglet scalar mesons $f_0(1710)$, $f_0(1500)$, $f_0(1370)$
and their mixing are studied. We employ two recent lattice results
as the starting point; one is the isovector scalar meson
$a_0(1450)$ which displays an unusual property of being nearly
independent of quark mass for quark masses smaller than that of
the strange, and the other is the scalar glueball mass at 1710 MeV
in the quenched approximation. In the SU(3) symmetry limit,
$f_0(1500)$ turns out to be a pure SU(3) octet and is degenerate
with $a_0(1450)$, while $f_0(1370)$ is mainly an SU(3) singlet
with a slight mixing with the scalar glueball which is the primary
component of $f_0(1710)$. These features remain essentially
unchanged even when SU(3) breaking is taken into account. We
discuss the sources of SU(3) breaking and their consequences on
flavor-dependent decays of these mesons. The observed enhancement
of $\omega f_0(1710)$ production over $\phi f_0(1710)$ in hadronic
$J/\psi$ decays and the copious $f_0(1710)$ production in
radiative $J/\psi$ decays lend further support to the prominent
glueball nature of $f_0(1710)$.

\vskip 7cm \noindent $^\dagger$ Address after August, 2006 :
Department of Physics, Chung Yuan Christian University, Chung-Li,
Taiwan 320, R.O.C.

\pagebreak

\section{Introduction}

Despite of the fact that the $q\bar{q}$ and glueball contents of
the iso-singlet scalar mesons $f_0(1710)$, $f_0(1500)$ and
$f_0(1370)$ have been studied extensively, it has been
controversial as to which of these is the dominant glueball.
Partly due to the fact that $f_0(1500)$, discovered in $p\bar{p}$
annihilation at LEAR, has decays to $\pi\pi,K\bar K$, $\eta\eta$
and $\eta\eta'$ modes which  are not compatible with a simple
$\bar qq$ picture \cite{ams95} and that the earlier quenched
lattice calculations~\cite{bsh93} predict the scalar glueball mass
to be $\sim 1550$ MeV, it has been suggested that $f_0(1500)$ is
primarily a scalar glueball~\cite{Close1}. Furthermore, because of
the small production of $\pi\pi$ in $f_0(1710)$ decay compared to
that of $K\bar K$, it has been thought that $f_0(1710)$ is
primarily $s\bar s$ dominated. On the other hand, the smaller
production rate of $K\bar K$ relative to $\pi\pi$ in $f_0(1370)$
decay leads to the conjecture that $f_0(1370)$ is governed by the
non-strange light quark content.

Based on the above observations, a flavor-mixing scheme is
proposed \cite{Close1} to consider the glueball and $q\bar q$
mixing in the neutral scalar mesons $f_0(1710)$, $f_0(1500)$ and
$f_0(1370)$. $\chi^2$ fits to the measured scalar meson masses and
their branching ratios of strong decays have been performed in
several references by Amsler, Close and Kirk~\cite{Close1}, Close
and Zhao~\cite{Close2}, and He {\it et al.}~\cite{He}. A common
feature of these analyses is that, before mixing, the $s\bar{s}$
mass $M_S$ is larger than the pure glueball mass $M_G$ which, in
turn, is larger than the $N(\equiv(u\bar{u}+d\bar{d})/\sqrt{2})$
mass $M_N$, with $M_G$ close to 1500 MeV and $M_S-M_N$ of the
order of $200\sim 300$ MeV. However, there are several serious
problems with this scenario. First, the isovector scalar meson
$a_0(1450)$ is confirmed to be a $q\bar{q}$ meson in lattice
calculations~\cite{Mathur,lw00,bde02,kun04,pdi04} which will be
discussed later. As such, the degeneracy of $a_0(1450)$ and
$K_0^*(1430)$, which has a strange quark, cannot be explained if
$M_S$ is larger than $M_N$ by $\sim 250$ MeV. Second, the most
recent quenched lattice calculation with improved action and
lattice spacings extrapolated to the continuum favors a larger
scalar glueball mass close to 1700 MeV~\cite{Chen,MP} (see below
for discussion). Third, if $f_0(1710)$ is dominated by the $s\bar
s$ content, the decay $J/\psi\to \phi f_0(1710)$ is expected to
have a rate larger than that of $J/\psi\to \omega f_0(1710)$.
Experimentally, it is other way around: the rate for $\omega
f_0(1710)$ production is about 6 times that of $J/\psi\to \phi
f_0(1710)$. Fourth, it is well known that the radiative decay
$J/\psi\to \gamma f_0$ is an ideal place to test the glueball
content of $f_0$. If $f_0(1500)$ has the largest scalar glueball
component, one expects  the $\Gamma(J/\psi\to \gamma f_0(1500))$
decay rate to be substantially larger than that of
$\Gamma(J/\psi\to \gamma f_0(1710))$. Again, experimentally, the
opposite is true.

Other scenarios have been proposed. Based on their lattice
calculations of the quenched scalar glueball mass at $1625(94)$
MeV at the infinite volume and continuum limits~\cite{wei94} and
the $s\bar{s}$ meson mass in the connected insertion (no
annihilation) at $\sim 1500$ MeV, Lee and
Weingarten~\cite{lw97,Lee} considered a mixing scheme where
$f_0(1500)$ is an almost pure $s\bar{s}$ meson and $f_0(1710)$ and
$f_0(1370)$ are primarily the glueball and $u\bar{u}+d\bar{d}$
meson respectively, but with substantial mixing between the two
($\sim 25\%$ for the small component). With the effective chiral
Lagrangian approach, Giacosa {\it et al.}~\cite{Giacosa} performed
a fit to the experimental masses and decay widths of $f_0(1710)$,
$f_0(1500)$ and $f_0(1370)$ and found four possible solutions,
depending on whether the direct decay of the glueball component is
considered. One of the solutions (see Appendix A) gives
$f_0(1710)$ as the pure glueball, while $f_0(1370)$ and
$f_0(1500)$ are dominated by the quarkonia components, but with
strong mixing between $(u\bar{u}+d\bar{d})/\sqrt{2}$ and $s\bar
s$. In this case, $M_S=1452$ MeV, $M_N=1392$ MeV and $M_G=1712$
MeV.

In this work, we shall employ two recent lattice results as the
input for the mass matrix which is essentially the starting point
for the mixing model between scalar mesons and the glueball. First
of all, an improved quenched lattice calculation of the glueball
spectrum at the infinite volume and continuum limits based on much
larger and finer lattices have been carried out~\cite{Chen}. The
mass of the scalar glueball is calculated to be
$m(0^{++})=1710\pm50\pm 80$ MeV. The implicit assumption of the
mixing models entails that the experimental glueball mass is
reflected as a result of mixing of the glueball in the quenched
approximation with the scalar $q\bar{q}$ mesons.  This suggests
that $M_G$ should be close to 1700 MeV rather than 1500
MeV~\cite{bsh93} from the earlier lattice calculations\footnote{We
should note that the Monte Carlo results of these two set of
calculations actually agree in lattice units within errors. The
difference comes from the fact that Ref.~\cite{bsh93} uses the
string tension to set the scale, while Ref.~\cite{Chen} uses the
Sommer scale~\cite{Sommer} $r_0=0.5$ fm to set the scale. It is
well-known that the scale of the quenched approximation is
uncertain from 10\% to 20\% depending on how it is set. The
relation between the string tension scale and the $r_0$ scale in
the quenched case has been examined for an extended range of
couplings (Wilson $\beta=6/g_0^2$ from 5.7 to 6.6)~\cite{gho00}
and it is found that they consistently differ by $\sim 10\%$ which
means that the string tension scale corresponds to $r_0 = 0.55$ fm
in the Sommer scale and thus places the predicted glueball mass at
1550 MeV, 10\% lower than the 1710 MeV set by the $r_0 = 0.5$ fm
scale. The inconsistency of scales in the quenched case has
prompted the work by the HPQCD-UKQCD-MILC-Fermilab collaboration
to study the issue in both the quenched approximation and in full
QCD~\cite{HUMF}. When a set of quantities from both the light
quark and heavy quark sectors are compared to experiments, they
found that the few outliers in the quenched approximation (e.g.
$f_{\pi}$ is higher than the average by $\sim 10\%$ and the
$1P-1S$ splitting in the Upsilon is lower than the average by
$\sim 10\%$) would line up with the rest to give a common scale in
full QCD. Since the string tension scale, like the $f_{\pi}$
scale, in the quenched approximation is 10\% higher than the
average while the latter is closer to that in full QCD, we think
it is essential to take $r_0=0.5$ fm and not the string tension to
set the scale in the quenched approximation in order to fairly
asses the quenched errors when compared with experiments.}.
  Second, the recent quenched lattice
calculation of the isovector scalar meson $a_0$ mass has been
carried out for a range of low quark masses~\cite{Mathur}. With
the lowest one corresponding to $m_{\pi}$ as low as 180 MeV, it is
found that, when the quark mass is smaller than that of the
strange, $a_0$ mass levels off, in contrast to those of $a_1$ and
other hadrons that have been calculated on the lattice. This
confirms the trend that has been observed in earlier works at
higher quark masses in both the quenched and unquenched
calculations~\cite{lw00,bde02,kun04,pdi04}. The chiral
extrapolated mass $a_0 = 1.42 \pm 0.13$ GeV suggests that
$a_0(1450)$ is a $q\bar{q}$ state. By virtue of the fact that
lower state $a_0(980)$ is not seen, it is concluded that
$a_0(980)$ is not a $q\bar{q}$ meson which is supposed to be
readily accessible with the $\overline{\psi}\psi$ interpolation
field. Furthermore, $K_0^{*}(1430)^+$, an $u\bar{s}$ meson, is
calculated to be $1.41 \pm 0.12$ GeV and the corresponding scalar
$\bar{s}s$ state from the connected insertion is $1.46 \pm 0.05$
GeV. This explains the fact that $K_0^{*}(1430)$ is basically
degenerate with $a_0(1450)$ despite having one strange quark. This
unusual behavior is not understood as far as we know and it serves
as a challenge to the existing hadronic models\footnote{There are
some attempts to understand the near degeneracy of $a_0(1450)$ and
$K^*_0(1430)$. Since the 4-quark light scalar nonet is known to
have a reversed ordering, namely, the scalar strange meson
$\kappa$ is lighter than the non-strange one such as $f_0(980)$,
it has been proposed in \cite{Schechter} to consider the mixing of
the $q\bar q$ heavy scalar nonet with the light nonet to make
$a_0(1450)$ and $K^*_0(1430)$ closer.  It goes further in
\cite{Maiani} to assume that the observed heavy scalar mesons form
another 4-quark nonet. We note, however, the quenched lattice
calculations in \cite{Mathur}, which presumably gives the bare
$q\bar{q}$ states before mixing with $q^2\bar{q}^2$ via sea quark
loops in the dynamical fermion calculation, already seem to
suggest the near degenercy between $a_0(1450)$ and
$K^*_0(1430)$.}. We are aware that there is a recent $N_f =2$
dynamical fermion calculation of $a_0$ with the
$\overline{\psi}\psi$ interpolation field~\cite{mm06}.
Extrapolating the mass difference between $b_0$ and $a_0$ to the
chiral limit, it is claimed that $a_0(980)$ is a $q\bar{q}$ state.
Even though there is no ghost state contamination in this
calculation as there are in the above-mentioned quenched and
partially quenched calculations, there is a physical $\pi\eta_2$
($\eta_2$ is the $\eta'$ in the $N_f=2$ case) nearby which has not
been taken into account. Also, since the quark masses in the
present calculation are heavy, it would not be able to discern the
possibility that $b_0$ and $a_0$ may cross each other toward the
chiral limit. In addition, it is known that there are a host of
problems assigning $a_0(980)$ to the $q\bar{q}$ state
phenomenologically. Here are some of them:

\begin{itemize}
\item

     If the $q\bar{q}$ $a_0$ indeed goes down with the quark mass as
is claimed in Ref.~\cite{mm06}, the same calculation with a
strange quark instead of an $u$ quark would yield an $s\bar{d}$
meson around $1100$ MeV which would place it far away from the two
known mesons in this mass range. In other words, it is $\sim 300$
MeV below $K_0^*(1430)$ and $\sim 300$ MeV above $\kappa(800)$.

\item

    It cannot explain why the $K_0^*(1430)$ which, according to the
review of scalar mesons in the particle data table, is a
$q\bar{q}$ state in all the models, is higher than the
axial-vector mesons $K_1(1270)$ and $K_1(1400)$. This is a
situation which parallels to the case of non-strange mesons where
$a_0(1450)$ is higher than $a_1(1260)$. The authors admitted this
is a problem in their paper~\cite{mm06}, but did not offer any
answer.

\item

 The widths of $a_0(980)$ and $f_0(980)$ are substantially
smaller than those of $a_0(1450)$ and $f_0(1370)$. In particular,
they are much smaller than that of $\kappa(800)$ which should be a
nonet partner with $a_0(980)$ and $f_0(980)$.

\item

    The $\gamma\gamma$ widths of $a_0(980)$ and $f_0(980)$ are much smaller
than expected of a $q\bar{q}$ state~\cite{bar85}.

\item

    It is hard to understand why $a_0(980)$ and $f_0(980)$ are basically degenerate.
The experimental data on $D_s^+ \rightarrow f_0(980)
\pi^+$~\cite{ait01} and $\phi \rightarrow
f_0(980)\gamma$~\cite{alo02} imply
   copious $f_0(980)$ production via its $s\bar{s}$ component. Yet, there cannot be
an $s\bar{s}$ component in $a_0(980)$ since it is an I=1 state.

\item The radiative decay $\phi\to a_0(980)\gamma$, which cannot
proceed if $a_0(980)$ is a $q\bar q$ state, can be nicely
described in the kaon loop mechanism \cite{Schechter06}. This
suggests a considerable admixture of the $K\bar K$ component.

\end{itemize}

   For all these reasons, we do not take the claim in
Ref.~\cite{mm06} seriously. We shall rely on the conclusion from
the other calculations in both the quenched and unquenched
calculations~\cite{Mathur,lw00,bde02,kun04,pdi04}.

As we discussed above, the accumulated lattice results hint at an
SU(3) symmetry in the scalar meson sector. Indeed, the near
degeneracy of $K_0^*(1430)$, $a_0(1470)$, and $f_0(1500)$ implies
that, to first order approximation, flavor SU(3) is a good
symmetry for the scalar mesons above 1 GeV, much better than the
pseudoscalar, vector, axial, and tensor sectors.

This work is organized as follows. In Sec. II, we discuss the
mixing matrix of scalar mesons in the SU(3) limit and its
implications on the strong decays of $f_0(1500)$. SU(3) breaking
effects and chiral suppression in the scalar glueball decay into
two pseudoscalar mesons are studied in Sec. III. Sec. IV is
devoted to the numerical results for the mixing matrix and
branching ratios. Conclusions are presented in Sec. V. The mixing
matrices of the isosinglet scalar mesons $f_0(1710)$, $f_0(1500)$
and $f_0(1370)$ that have been proposed in the literature are
summarized in the appendix.

\section{Mixing matrix}
 We shall use $|U\ra, |D\ra, |S\ra$ to denote the
quarkonium states $|u\bar u\ra, |d\bar d\ra$ and $|s\bar s\ra$,
and $|G\ra$ to denote the pure scalar glueball state. In this
basis, the mass matrix reads
 \be \label{eq:massmatrix}
 {\rm M}=\left( \matrix{ M_U & 0 & 0 & 0  \cr
                   0 & M_D & 0 & 0  \cr
                    0 & 0 & M_S & 0  \cr
                      0 & 0 & 0 & M_G  \cr
                  } \right)+\left(\matrix{ x & x & x_s & y \cr
                x & x & x_s & y \cr
                x_s & x_s & x_{ss} & y_s \cr
                y & y & y_s & 0 \cr} \right),
 \en
where the mass parameters $m_G$ is the mass of the scalar glueball
in the pure gauge sector, and $m_{U,D,S}$ are the scalar quarkonia
$u\bar u$, $d\bar d$ and $s\bar s$ before mixing which correspond
to those from the connected insertion calculation. The parameter
$x$ denotes the mixing between different $q\bar{q}$ states through
quark-antiquark annihilation and $y$ stands for the
glueball-quarkonia mixing strength. Possible SU(3) breaking
effects are characterized by the subscripts ``$s$" and ``$ss$". As
noticed in passing, lattice calculations~\cite{Mathur} of the
$a_0(1450)$ and $K_0^*(1430)$ masses indicate a good SU(3)
symmetry for the scalar meson sector above 1 GeV. This means that
$M_S$ should be close to $M_U$ or $M_D$. Also the glueball mass
$m_G$ should be close to the scalar glueball mass $1710\pm50\pm80$
MeV from the lattice QCD calculation in the pure gauge sector
\cite{Chen}.

We shall begin by considering exact SU(3) symmetry as a first
approximation, namely, $M_S=M_U=M_D=M$ and $x_s=x_{ss}=x$ and
$y_s=y$. In this case, it is convenient to recast the mass matrix
in Eq. (\ref{eq:massmatrix}) in terms of the basis
$|a_0(1450)\ra$, $|f_{\rm octet}\ra$, $|f_{\rm singlet}\ra$ and
$|G\ra$ defined by
 \be
&& |a_0(1450)\ra={1\over\sqrt{2}}(|u\bar u\ra-|d\bar d\ra), \non\\
  && |f_{\rm octet}\ra= {1\over\sqrt{6}}(|u\bar u\ra+|d\bar d\ra-2|s\bar
  s\ra), \qquad\quad |f_{\rm singlet}\ra={1\over\sqrt{3}}(|u\bar u\ra+|d\bar d\ra+|s\bar
  s\ra).
  \en
Then the mass matrix becomes
 \be \label{eq:massmatrix1}
 {\rm M}=\left( \matrix{ M & 0 & 0 & 0  \cr
                   0 & M & 0 & 0  \cr
                    0 & 0 & M+3x & \sqrt{3}y  \cr
                      0 & 0 & \sqrt{3}y & M_G  \cr
                  } \right).
                  \en
The first two eigenstates are identified with $a_0(1450)$ and
$f_0(1500)$ which are degenerate with the mass $M$. Taking $M$ to
be the experimental mass of $1474\pm 19$ MeV \cite{PDG}, we see it
is a good approximation for the mass of $f_0(1500)$ at $1507\pm 5$
MeV \cite{PDG}. Thus, in the limit of exact SU(3) symmetry,
$f_0(1500)$ is the SU(3) isosinglet octet state $|f_{\rm
octet}\ra$ and is degenerate with $a_0(1450)$. The diagonalization
of the lower $2\times 2$ matrix in (\ref{eq:massmatrix1}) yields
the eigenvalues
 \be
 m_{f_0(1370)}=\ov M-\sqrt{\Delta^2+3y^2}, \qquad  m_{f_0(1700)}=\ov M+\sqrt{\Delta^2+3y^2},
 \en
where $\ov M=(M+3x+M_G)/2$ and $\Delta=(M_G-M-3x)/2$, and the
corresponding eigenvectors are
 \be \label{eq:1370&1700}
 |f_0(1370)\ra &=& N_{1370}\left(|f_{\rm singlet}\ra-{\sqrt{3}y\over
 \Delta+\sqrt{\Delta^2+3y^2}}|G\ra \right), \non \\
 |f_0(1710)\ra &=&
 N_{1710}\left(|G\ra+{\sqrt{3}y\over
 \Delta+\sqrt{\Delta^2+3y^2}}|f_{\rm
 singlet}\ra \right),
 \en
with $N_{1370}$ and $N_{1700}$ being the normalization constants.

Several remarks are in order. (i) In the absence of
glueball-quarkonium mixing, i.e. $y=0$, we see from Eq.
(\ref{eq:1370&1700}) that $f_0(1370)$ becomes a pure SU(3) singlet
$|f_{\rm singlet}\ra$ and $f_0(1710)$ the pure glueball $|G\ra$.
The $f_0(1370)$ mass is given by $m_{f_0(1370)}=M+3x$. Taking the
experimental $f_0(1370)$ mass to be $1370$ MeV, the
quark-antiquark mixing matrix element $x$ through annihilation is
found to be $-33$ MeV. (ii) When the glueball-quarkonium mixing
$y$ is turned on, there will be some mixing between the glueball
and the SU(3)-singlet $q\bar{q}$ . If $y$ has the same magnitude
as $x$, i.e. $33$ MeV, then $3y^2 \ll \Delta^2$ where $\Delta$ is
half of the mass difference between $M_G$ and $M+3x$, which is
$\sim 170$ MeV. In this case, the mass shift of $f_0(1370)$ and
$f_0(1710)$ due to mixing is only $\sim 3y^2/2\Delta = 9.6$ MeV.
In the wavefunctions of the mixed states, the coefficient of the
minor component is of order $\sqrt{3}y/(2\Delta) = 0.17$ which
corresponds to $\sim 3\%$ mixing.

We next proceed to consider the implications of the aforementioned
mixing scheme to strong decays. We first discuss the $f_0(1500)$
meson, since its strong decays are better measured. If $f_0(1500)$
is the octet state $|f_{\rm octet}\ra$, it will lead to the
predictions
 \be
 {\Gamma(f_0(1500)\to K\bar K)\over\Gamma(f_0(1500)\to
 \pi\pi)}\approx 0.21, \qquad  {\Gamma(f_0(1500)\to \eta\eta)\over\Gamma(f_0(1500)\to
 \pi\pi)}\approx 0.02,
 \en
where we have used the $\eta-\eta'$ mixing angle
$\theta=-(15.4\pm1.0)^\circ$ \cite{Kroll}. The corresponding
experimental results are $0.246\pm0.026$ and $0.145\pm0.027$
\cite{PDG}. We see that although the ratio of $K\bar K/\pi\pi$ is
well accommodated, the predicted ratio for $\eta\eta/\pi\pi$ is
too small. This can be understood as follows. Assuming no SU(3)
breaking in the decay amplitude, we obtain \footnote{Eq.
(\ref{eq:1500decay}) also can be obtained from Eq.
(\ref{eq:ampsquare}) by neglecting the glueball contribution from
$f_0(1500)$.}
 \be \label{eq:1500decay}
 {\Gamma(f_0(1500)\to K\bar K)\over\Gamma(f_0(1500)\to
 \pi\pi)} &=& {1\over 3}\left(1+{s_2\over u_2}\right)^2{p_K\over
 p_\pi}, \non \\
  {\Gamma(f_0(1500)\to \eta\eta)\over\Gamma(f_0(1500)\to
 \pi\pi)} &=& {1\over 27}\left(2+{s_2\over u_2}\right)^2{p_\eta\over
 p_\pi},
 \en
where $p_h$ is the c.m. momentum of the hadron $h$, $u_2$ and
$s_2$ are the $f_0(1500)$ wavefunction coefficients defined in the
orthogonal transformation $U$
 \be
 \left(\matrix{ a_0(1450) \cr f_0(1500) \cr f_0(1370) \cr f_0(1710) \cr}\right)=U\left(\matrix{  |U\ra \cr
 |D\ra  \cr  |S\ra \cr |G\ra \cr}\right)=
\left( \matrix{ u_1 & d_1 & s_1 & g_1 \cr
                u_2 & d_2 & s_2 & g_2 \cr
                u_3 & d_3 & s_3 & g_3 \cr
                 u_4 & d_4 & s_4 & g_4 \cr
                  } \right)\left(\matrix{|U\ra \cr |D\ra \cr
 |S\ra \cr |G\ra \cr}\right).
 \en
For $a_0(1450)$, $s_1=g_1=0$ and $u_1=-d_1={1\over\sqrt{2}}$. To
derive Eq. (\ref{eq:1500decay}) we have, for simplicity, applied
the $\eta-\eta'$ mixing angle $\theta=-19.5^\circ$ so that the
wave functions of $\eta$ and $\eta'$ have the simple expressions:
 \be
 |\eta\ra= {1\over \sqrt{3}}|u\bar u+d\bar d-s\bar s\ra, \qquad
 |\eta'\ra={1\over\sqrt{6}}|u\bar u+d\bar d+2s\bar s\ra.
 \en
Since $s_2= -2u_2=-2d_2$ for the octet $f_0(1500)$, it is evident
from Eq. (\ref{eq:1500decay}) that $f_0(1500)$ will not decay into
$\eta\eta$ (or strongly suppressed) if SU(3) symmetry is exact.
This implies that SU(3) symmetry must be broken in the mass matrix
and/or in the decay amplitudes.

\section{SU(3) breaking and Chiral suppression}
 As discussed before, SU(3) symmetry leads naturally to the near
degeneracy of $a_0(1450)$, $K_0^*(1430)$ and $f_0(1500)$. However,
in order to accommodate the observed branching ratios of strong
decays, SU(3) symmetry must be broken to certain degree in the
mass matrix and/or in the decay amplitudes. One also needs
$M_S>M_U=M_D$ in order to lift the degeneracy of $a_0(1450)$ and
$f_0(1500)$. Since the SU(3) breaking effect is expected to be
weak, they will be treated perturbatively. In the mass matrix in
Eq. (\ref{eq:massmatrix}), the glueball-quarkonia mixing has been
computed in lattice QCD  with the results \cite{Lee}
 \be   \label{eq:LQCDy}
 y=43\pm31\,{\rm MeV}, \qquad y/y_s=1.198\pm 0.072\,,
 \en
which confirms that the magnitudes of $y$ and $x$ are about the
same, as expected.

For strong decays, we consider a simple effective Hamiltonian of a
scalar state decaying into two pseudoscalar mesons for the OZI
allowed, OZI suppressed, and doubly OZI suppressed
interactions:\footnote{In principle, one can add more interaction
terms such as ${\rm Tr}[X_F]{\rm Tr}[PP]$ terms (see
 \cite{He} for detail). Due to the presumed narrowness of the glueball width,
 we assume its decay to hadrons is large $N_c$ suppressed. We still call
 them OZI suppressed and doubly OZI suppressed in terms of the ways the mesons are
 formed from the quark lines. }
 \be \label{eq:Hamiltonian}
 {\cal H}_{SPP}=f_1{\rm Tr}[X_FPP]+f_2X_G{\rm Tr}[PP]+f_3X_G{\rm
 Tr}[P]{\rm Tr}[P],
 \en
where
 \be
 X_F=\left(\matrix{ u\bar u & 0 & 0 \cr
                  0 & d\bar d & 0 \cr
                  0 & 0 & s\bar s \cr}\right)=
   \left(\matrix{ \sum u_iF_i & 0 & 0 \cr
                  0 & \sum d_iF_i & 0 \cr
                  0 & 0 & \sum s_iF_i \cr}\right), \qquad
                  X_G=\sum g_iF_i,
                  \en
with $F_i=(a_0(1450),f_0(1500),f_0(1370),f_0(1710))$. $P$ is the
pseudoscalar nonet
 \be
 P &=& \left(\matrix{{\pi^0\over\sqrt{2}}+{\eta_8\over\sqrt{6}}+{\eta_0\over\sqrt{3}}
 & \pi^+ & K^+ \cr
 \pi^- & -{\pi^0\over\sqrt{2}}+{\eta_8\over\sqrt{6}}+{\eta_0\over\sqrt{3}}
 & K^0 \cr
 K^- & \ov K^0 & -{2\eta_8\over\sqrt{6}}+{\eta_0\over\sqrt{3}}
 \cr}\right) \non \\
 &=&
 \left(\matrix{{\pi^0\over\sqrt{2}}+{a_\eta\eta+a_{\eta'}\eta'\over\sqrt{2}}
 & \pi^+ & K^+ \cr
 \pi^- & -{\pi^0\over\sqrt{2}}+{a_\eta\eta+a_{\eta'}\eta'\over\sqrt{2}}
 & K^0 \cr
 K^- & \ov K^0 & b_\eta\eta+b_{\eta'}\eta'
 \cr}\right),
 \en
with
 \be
 && a_\eta=b_{\eta'}={\cos\theta-\sqrt{2}\sin\theta\over\sqrt{3}},
 \non \\
 &&
 a_{\eta'}=-b_{\eta}={\sin\theta+\sqrt{2}\cos\theta\over\sqrt{3}}.
 \en
In the above equation, $\theta$ is the $\eta-\eta'$ mixing angle
defined by
 \be
 \eta=\eta_8\cos\theta-\eta_0\sin\theta,\qquad\quad
 \eta'=\eta_8\sin\theta+\eta_0\cos\theta.
 \en

The invariant amplitudes squared for various strong decays are
given by
 \be \label{eq:ampsquare}
 |A(F_i\to K\bar K)|^2 &=& 2f_1^2(r_au_i+s_i+2\rho_s^{KK}g_i)^2, \non \\
 |A(F_i\to \pi\pi)|^2 &=& 6f_1^2(u_i+\rho_s^{\pi\pi}g_i)^2, \non \\
 |A(F_i\to \eta\eta)|^2 &=&
 2f_1^2\left(a_\eta^2u_i+r_ab_\eta^2s_i+\rho_s^{\eta\eta}(a_\eta^2
 +b_\eta^2)g_i+\rho_{ss}(2a_\eta^2+b_\eta^2+{4\over\sqrt{2}}a_\eta
 b_\eta)g_i\right)^2,  \\
 |A(a_0\to \pi\eta)|^2 &=&  4f_1^2a_\eta^2u_1^2, \non
 \en
where $\rho_s=f_2/f_1$ and $\rho_{ss}=f_3/f_1$ are the ratios of
the OZI suppressed and the doubly OZI suppressed couplings to that
of the OZI allowed one. The invariant amplitudes squared for
$f_2(1270)\to K\bar K,\pi\pi,\eta\eta$ are similar to that of
$F_i$. (The quark content of $f_2(1270)$ is $(u\bar u+d\bar
d)/\sqrt{2}$.) Likewise, the invariant amplitudes squared for
$a_2(1320)\to K\bar K$ and $\pi\eta$ are similar to that of
$a_0(1450)$. The contribution characterized by the coupling $f_3$
or $\rho_{ss}$ arises only from the SU(3)-singlet $\eta_0$. The
parameter $r_a$ denotes a possible SU(3) breaking effect in the
OZI allowed decays when the $s\bar s$ pair is created relative to
the $u\bar{u}$ and $d\bar{d}$ pairs. In principle, it can be
determined from the ratios ${\Gamma(a_0(1450)\to K\bar K)\over
\Gamma(a_0(1450)\to \pi\eta)}$, ${\Gamma(a_2(1320)\to K\bar
K)\over \Gamma(a_2(1320)\to \pi\eta)}$,
$\frac{\Gamma(f_2(1270)\rightarrow K\bar
K)}{\Gamma(f_{2}(1270)\rightarrow \pi\pi)}$ and
$\frac{\Gamma(f_2(1270)\rightarrow \eta\eta)}
{\Gamma(f_{2}(1270)\rightarrow \pi\pi)}$.

We now explain why we put the superscripts $K\bar{K}$, $\pi\pi$
and $\eta\eta$ to the parameter $\rho_s$ in Eq.
(\ref{eq:ampsquare}). It is clear from Eq. (\ref{eq:ampsquare})
that, for a pure glueball decay into $\pi\pi$ and $K\bar K$, we
have
 \be
 {\Gamma(G\to \pi\pi)\over\Gamma(G\to K\bar K)}={3\over
 4}\left({\rho_s^{\pi\pi}\over \rho_s^{K\bar{K}}}\right)^2{p_\pi\over p_K}.
 \en
If the coupling $f_2$ (and hence $\rho_s$) between glueball and
two pseudoscalar mesons is flavor independent, i.e.
$\rho_s^{K\bar{K}}=\rho_s^{\pi\pi}$, then it is expected that
$\Gamma(G\to \pi\pi)/\Gamma(G\to K\bar K)=0.91$. Since
$\Gamma(f_0(1710)\to \pi\pi)/\Gamma(f_0(1710)\to K\bar K)$ is
measured to be $0.20\pm0.04$ by WA102 \cite{Barberis}, $<0.11$ by
BES from $J/\psi\to\omega(K^+K^-,\pi^+\pi^-)$ decays
\cite{BESomegaKK} and $0.41^{+0.11}_{-0.17}$ from
$J/\psi\to\gamma(K^+K^-,\pi^+\pi^-)$ decays
\cite{BESgammapipi},\footnote{For the purpose of fitting, we will
use $\Gamma(f_0(1710)\to \pi\pi)/\Gamma(f_0(1710)\to K\bar
K)=0.30\pm0.20$ as the experimental input.}
this implies a relatively large suppression of $\pi\pi$ production
relative to $K\bar K$ in scalar glueball decays if $f_0(1710)$ is
a pure $0^{++}$ glueball. To explain the large disparity between
$\pi\pi$ and $K\bar K$ production in scalar glueball decays,
Chanowitz \cite{Chanowitz} advocated that a pure scalar glueball
cannot decay into quark-antiquark in the chiral limit, i.e.
\begin{equation}  \label{eq:chiral_supp}
 A(G\to q\bar q)\propto m_q.
 \end{equation}
Since the current strange quark mass is an order of magnitude
larger than $m_u$ and $m_d$, decay to $K\bar K$ is largely favored
over $\pi\pi$. Furthermore, it has been pointed out that chiral
suppression will manifest itself at the hadron level~\cite{Chao}.
To this end, it is suggested in~\cite{Chao} that $m_q$ in Eq.
(\ref{eq:chiral_supp}) should be interpreted as the scale of
chiral symmetry breaking since chiral symmetry is broken not only
by finite quark masses but is also broken spontaneously.
Consequently, chiral suppression for the ratio $\Gamma(G\to
\pi\pi)/\Gamma(G\to K\bar K)$ is not so strong as the current
quark mass ratio $m_u/m_s$. A pQCD calculation in \cite{Chao}
yields
 \be
 {A(G\to \pi^+\pi^-)\over A(G\to K^+K^-)}\approx \left({f_\pi\over
f_K}\right)^2,
 \en
due mainly to the difference of the $\pi$ and $K$ light-cone
distribution functions. Lattice calculations \cite{Sexton} seem to
confirm the chiral suppression effect (see footnote 2 of
\cite{Burakovsky}) with the results
 \be \label{eq:c1flavor}
 \rho_s^{\pi\pi}:\rho_s^{K\bar
 K}:\rho_s^{\eta\eta}=0.834^{+0.603}_{-0.579}:2.654^{+0.372}_{-0.402}:3.099^{+0.364}_{-0.423}\,.
 \en
which are in sharp contrast to the flavor-symmetry limit with
$\rho_s^{\pi\pi}:\rho_s^{K\bar K}:\rho_s^{\eta\eta}=1:1:1$.
Although the errors are large, the lattice results show a sizable
deviation from this limit.

>From Eq. (\ref{eq:ampsquare}), the ratio of $\pi\pi$ and $K\bar K$
productions in $f_0(1710)$ decays is given by
 \be \label{eq:f1710topipiKK}
 {\Gamma(f_0(1710)\to \pi\pi)\over\Gamma(f_0(1710)\to K\bar K)}=
 3\left({u_4+\rho_s^{\pi\pi}g_4\over r_au_4+s_4+2\rho_s^{K\bar{K}}g_4}\right)^2\,{p_\pi\over p_K}.
 \en
At first sight, it appears that if $\rho_s$ is negative, the
destructive interference between the glueball and quark contents
of $f_0(1710)$ may lead to the desired suppression of $\pi\pi$
production even if the glueball decay is flavor blind, i.e.
$\rho_s^{K\bar K}=\rho_s^{\pi\pi}$. That is, it seems possible
that one does not need chiral suppression in order to explain the
observed suppression of $\pi\pi$ relative to $K\bar K$. However,
as we shall see below, chiral non-suppression will lead to too
small a width for $f_0(1710)$ which is several orders of magnitude
smaller than the experimental width of $138\pm9$ MeV \cite{PDG}.
As stressed in \cite{Giacosa}, it is important to impose the
condition that the total sum of the partial decay widths of
$f_0(1710)$ into two pseudoscalar mesons to be comparable to but
smaller than the total width. Without such a constraint, a local
minimum for $\chi^2$ can occur where $\Gamma[f_0(1710)]$ is either
too large or too small compared to experiments.

\section{Numerical results}
 To illustrate our mixing model with numerical results, we first
fix the the parameters $x=-44$ MeV, $x_{ss}/x_s=x_s/x=0.82$ and
$\theta=-14.4^\circ$. For the chiral suppression in scalar
glueball decay, we parametrize $\rho_s^{\pi\pi}=\rho_s$,
$\rho_s^{K\bar{K}}=\rho_s r_s$ and
$\rho_s^{\eta\eta}(a_\eta^2+b_\eta^2)=\rho_s(a_\eta^2+r_s^2
b_\eta^2)$ with $r_s$ being an SU(3) breaking parameter in the OZI
suppressed decays.
 We shall consider two cases of chiral suppression; (i) $r_s=1.55$, so
that  $\rho_s^{\pi\pi}:\rho_s^{K\bar K}:\rho_s
^{\eta\eta}=1:1.55:1.59$ and (ii) $r_s=3.15$ which corresponds to
 $\rho_s^{\pi\pi}:\rho_s^{K\bar K}:\rho_s
^{\eta\eta}=1:3.15:4.74$. They are in the region allowed by Eq.
(\ref{eq:c1flavor}). Choosing $y=64$ MeV and $y/y_s=1.19$ in the
range constrained by Eq. (\ref{eq:LQCDy}) and performing a best
$\chi^2$ fit to the experimental masses of $f_0(1710)$,
$f_0(1500)$, $f_0(1370)$ and branching ratios of $f_0(1710)$,
$f_0(1500)$, $a_0(1450)$, $a_2(1320)$ and $f_2(1270)$, we obtain
the fitted parameters as shown in Table \ref{tab:fitparameter}. As
noticed in passing, the quarkonium $n\bar n$ mass $M_N$ is fixed
by the $a_0(1450)$ state. The fitted masses and branching ratios
are summarized in Table \ref{tab:fit}, while the predicted decay
properties of scalar mesons are exhibited in Table
\ref{tab:prediction}.

\begin{table}[t]
\caption{Fitted parameters for two cases of chiral suppression in
$G\to PP$ decay: (i) $r_s=1.55$ and (ii) $r_s=3.15$. Parameters
denoted by ``*" are input ones. } \label{tab:fitparameter}
\begin{ruledtabular}
\begin{tabular}{c c c c c c c c c}
& $M_N$ (MeV)$^*$ & $M_S$ (MeV)& $M_G$ (MeV) & $x_s/x^{*}$ & $y_s/y^*$ & $r_a$ & $\rho_s$ & $\rho_{ss}$ \\
\hline
 (i) & 1474 & 1507 & 1665 & 0.84 & 0.82 & 1.21  & $-0.48$ & 0 \\
 (ii) & 1474 & 1498 & 1666 & 0.84 & 0.82 & 1.22  & 0.10 & 0.12 \\
\end{tabular}
\end{ruledtabular}
\end{table}

Some of the strong decay modes are not used for the fit. The
experimental measurements of $\Gamma(f_0(1370)\to K\bar
K)/\Gamma(f_0(1370)\to\pi\pi)$ range from $1.33\pm0.67$
\cite{Bugg}, $0.91\pm0.20$ \cite{Bargiotti}, $0.46\pm0.15\pm0.11$
\cite{Barberis00} to $0.12\pm0.06$ \cite{Anisovich} and
$0.08\pm0.08$ \cite{BESphi}. Likewise, the result for
$f_0(1370)\to\eta\eta$ spans a large range. Consequently, the
decays of $f_0(1370)$ are not employed as the fitting input. The
decay $f_0(1500)\to \eta\eta'$ is also not used for the fit since
$\eta\eta'$ is produced at threshold and hence its measurement
could be subject to large uncertainties.

The mixing matrices obtained in both cases have the similar
results:
 \be \label{eq:wf}
 \left(\matrix{ f_0(1370) \cr f_0(1500) \cr f_0(1710) \cr}\right)=
\left( \matrix{ 0.78 & 0.51 & -0.36 \cr
                 -0.54 & 0.84 & 0.03 \cr
                0.32 & 0.18 & 0.93 \cr
                  } \right)\left(\matrix{|N\ra \cr
 |S\ra \cr |G\ra \cr}\right).
 \en
It is evident that $f_0(1710)$ is composed primarily of the scalar
glueball, $f_0(1500)$ is close to an SU(3) octet, and $f_0(1370)$
consists of an approximated SU(3) singlet with some glueball
component ($\sim 10\%$). Unlike $f_0(1370)$, the glueball content
of $f_0(1500)$ is very tiny because an SU(3) octet does not mix
with the scalar glueball.

\begin{table}[t]
\caption{Fitted masses and branching ratios for two cases of
chiral suppression in $G\to PP$ decay: (i) $r_s=1.55$
 and (ii) $r_s=3.25$. } \label{tab:fit}
\begin{ruledtabular}
\begin{tabular}{|c|cc c|}
&~~~~Experiment~~~~ & ~~fit (i)~~ & ~~fit (ii)~~ \\
\hline
 ~~$M_{f_{0}(1710)}$(MeV)~~&$1714\pm5$ \cite{PDG}&1715 & 1715 \\\hline
$M_{f_{0}(1500)}$(MeV)&$1507\pm5$ \cite{PDG}&1510 & 1504
\\ \hline $M_{f_{0}(1370)}$(MeV)&$1350\pm 150$ \cite{PDG}&1348 &
1346  \\ \hline
 $\frac{\Gamma(f_{0}(1500)\rightarrow
\eta\eta)}{\Gamma(f_{0}(1500)\rightarrow \pi\pi)}$&$0.145\pm0.027$ \cite{PDG} & 0.068 & 0.081 \\
 \hline $\frac{\Gamma(f_{0}(1500)\rightarrow
K\bar{K})}{\Gamma(f_{0}(1500)\rightarrow \pi\pi)}$& $0.246\pm0.026$ \cite{PDG} &0.26 & 0.27 \\
 \hline $\frac{\Gamma(f_{0}(1710)\rightarrow
\pi\pi)}{\Gamma(f_{0}(1710)\rightarrow K\bar{K})}$& $0.30\pm0.20$(see text)  &0.21 & 0.34 \\
 \hline $\frac{\Gamma(f_{0}(1710)\rightarrow
\eta\eta)}{\Gamma(f_{0}(1710)\rightarrow K\bar{K})}$&
$0.48\pm0.15$ \cite{Barberis} &0.26 & 0.51 \\
 \hline  $\frac{\Gamma(a_{0}(1450)\rightarrow
K\bar K)}{\Gamma(a_{0}(1450)\rightarrow \pi\eta)}$&
$0.88\pm0.23$ \cite{PDG} &1.10 & 1.12 \\
 \hline  $\frac{\Gamma(a_{2}(1320)\rightarrow
K\bar K)}{\Gamma(a_{2}(1320)\rightarrow \pi\eta)}$&
$0.34\pm0.06$ \cite{PDG} & 0.45 & 0.46 \\
 \hline $\frac{\Gamma(f_2(1270)\rightarrow K\bar
K)}{\Gamma(f_{2}(1270)\rightarrow \pi\pi)}$ &
$0.054^{+0.005}_{-0.006}$ \cite{PDG}  &0.056 & 0.057 \\
 \hline $\frac{\Gamma(f_2(1270)\rightarrow \eta\eta)}
{\Gamma(f_{2}(1270)\rightarrow \pi\pi)}$ &
$0.003\pm0.001$ \cite{Barberis00} &0.005 & 0.005 \\
 \hline $\chi^2/{\rm d.o.f.}$ &  & 2.6 & 2.5 \\
\end{tabular}
\end{ruledtabular}
\end{table}

\begin{table}[h]
\caption{Predicted decay properties of scalar mesons for  (i)
$r_s=1.55$ and (ii) $r_s=3.25$. For partial widths of $f_0(1710)$
and $f_0(1500)$, we have summed over $PP=K\bar K,\pi\pi,\eta\eta$
states.} \label{tab:prediction}
\begin{ruledtabular}
\begin{tabular}{|c|cc c|}
&~~~Experiment~~~ & ~~fit (i)~~ & ~~fit (ii)~~ \\ \hline
 $\frac{\Gamma(f_{0}(1370)\rightarrow
K\bar{K})}{\Gamma(f_{0}(1370)\rightarrow\pi\pi )}$& see text & 1.27 & 0.79 \\
\hline $\frac{\Gamma(f_{0}(1370)\rightarrow
\eta\eta)}{\Gamma(f_{0}(1370)\rightarrow K\bar{K})}$& $0.35\pm0.30$ \cite{PDG} & 0.21 & 0.12\\
\hline  $\frac{\Gamma(J/\psi\rightarrow \omega f_0(1710))}
{\Gamma(J/\psi\rightarrow \phi f_0(1710))}$& $6.6\pm2.7$
\cite{BESphi,BESomegaKK,Zhao}
 &  3.8 & 4.1 \\
 \hline $\frac{\Gamma(J/\psi\rightarrow \omega f_0(1500))}
{\Gamma(J/\psi\rightarrow \phi f_0(1500))}$&
$$ & 0.44 & 0.47 \\
 \hline $\frac{\Gamma(J/\psi\rightarrow \omega f_0(1370))}
{\Gamma(J/\psi\rightarrow \phi f_0(1370))}$&
$$ & 2.85 &  2.56 \\ \hline
 $\Gamma_{f_0(1710)\to PP}$(MeV) & $<138\pm9$ \cite{PDG} & 84 & 133 \\ \hline
  $\Gamma_{f_0(1370)\to PP}$(MeV) &  & 406 & 146 \\
\end{tabular}
\end{ruledtabular}
\end{table}

To compute the partial decay widths of $f_0(1710)$ and $f_0(1370)$
we have used the measured $\Gamma(f_0(1500)\to\pi\pi)=34\pm4$
MeV~\cite{PDG} to fix the strong coupling $f_1$. We see from Table
\ref{tab:prediction} that the predicted 2-body $PP$ decay width of
$f_0(1710)$ in case (i) is smaller than that in case (ii). This is
because, given a smaller $r_s (=1.55)$, the parameter $\rho_s$ has
to be negative in order to fit the ratio of
$\Gamma(\pi\pi)/\Gamma(K\bar K)$ in $f_0(1710)$ decay. This in
turn leads to a suppressed $f_0(1710)$ width due to the
destructive interference between the glueball and quarkonia
contributions. Note that, apart from the two-body decay modes
$K\bar K$, $\pi\pi$ and $\eta\eta$, none of the multihadron modes
in $f_0(1710)$ decay has been seen, though theoretically $G\to
q\bar q g$ and $G\to qq\bar q\bar q$ are not chirally suppressed
\cite{Chanowitz}. In our favored scenario (ii), the calculated
partial width of 133 MeV is in agreement with the scenario that
the decay of $f_0(1710)$ is saturated by the $PP$ pairs.

It should be stressed that, in the absence of chiral suppression
in $G\to PP$ decay; namely, $\rho_s^{\pi\pi}=\rho_s^{K\bar K}$,
the $f_0(1710)$ width is predicted to be less than 1 MeV and hence
is ruled out by experiment. This is a strong indication in favor
of chiral suppression of $G\to\pi\pi$ relative to $G\to K\bar K$.
We note that fitted $|\rho_s|$ and $|\rho_{ss}|$ are less than
unity in both (i) and (ii). This supports the supposition that the
OZI suppressed decays via the $f_2$ and $f_3$ terms in Eq.
(\ref{eq:Hamiltonian}) are smaller than the OZI allowed decay via
the $f_1$ term.

Apart from the partial widths of $f_0(1710)$ and $f_0(1370)$,
scenarios (i) and (ii) also differ in the predictions of
$f_0(1370)\to K\bar K/\pi\pi$ (see Table \ref{tab:prediction}) and
$f_0(1710)\to\eta\eta/K\bar K$ (Table \ref{tab:fit}). Because the
glueball and quark contents in the wavefunction of $f_0(1370)$
have an opposite sign and the parameter $\rho_s$ is negative in
the case of (i) as noted in passing, the interference between
$q\bar q$ and glueball amplitudes turns out to be constructive in
$f_0(1370)\to (\pi\pi, K\bar K)$ decays [see Eq.
(\ref{eq:f1710topipiKK}) with $f_0(1700)$ replaced by
$f_0(1370)$]. Consequently, the ratio $R\equiv \Gamma(f_0(1370)\to
K\bar K)/\Gamma(f_0(1370)\to \pi\pi)$ is larger than unity in case
(i). In our favored case (ii), $R$ is predicted to be 0.79\,. The
$\chi^2$ value in both cases is almost entirely governed by the
ratio ${\Gamma(f_0(1500)\to\eta\eta)\over
\Gamma(f_0(1500)\to\pi\pi)}$ whose measurement ranges from $
0.230\pm0.097$ \cite{Amsler95} to $0.18\pm0.03$ \cite{Barberis00}
and $0.080\pm0.033$ \cite{Amsler02}. If the measured ratio is
close to our prediction of 0.08, the $\chi^2$ value will be
greatly reduced.

In our scheme, it is easy to understand why $J/\psi\to\omega
f_0(1710)$ has a rate larger than $J/\psi\to \phi f_0(1710)$ This
is because the $n\bar n$ content is more copious than $s\bar s$ in
$f_0(1710)$. Just as the scalar meson decay into two pseudoscalar
mesons, we can use the similar Hamiltonian for the
vector-vector-scalar interaction as in Eq. (\ref{eq:Hamiltonian})
 \be \label{eq:Hamiltonian}
 {\cal H}_{S(J/\psi) V}=h_1{\rm Tr}[X_FV]+h_2X_G{\rm
 Tr}[V]+h_3{\rm Tr}[X_F]{\rm Tr}[V]
 \en
to write
 \be
 |A(J/\psi\to \phi F_i)|^2 = h_1^2s_i^2,  \qquad
 |A(J/\psi\to \omega F_i)|^2 &=& 2h_1^2u_i^2,
 \en
where we have neglected the $h_2$ and $h_3$ terms which are
presumably OZI  suppressed. Our prediction of $\Gamma(J/\psi\to
\omega f_0(1710))/\Gamma(J/\psi\to \phi f_0(1710))=4.1$ is
consistent with the observed value of $6.6\pm2.7$.\footnote{The
published BES measurements are $\B(J/\psi\to\phi f_0(1710)\to\phi
K\bar K)=(2.0\pm0.7)\times 10^{-4}$ \cite{BESphi} and
$\B(J/\psi\to\omega f_0(1710)\to\omega K\bar K)=(6.6\pm1.3)\times
10^{-4}$ \cite{BESomegaKK}. For the latter, we shall use the
updated value of $(13.2\pm2.6)\times 10^{-4}$ \cite{Zhao}.} If
$f_0(1710)$ is dominated by $s\bar s$ as advocated before
\cite{Close1,Close2}, one will naively expect a suppression of the
$\omega f_0(1710)$ production relative to $\phi f_0(1701)$. One
way to circumvent this apparent contradiction with experiment is
to assume a large OZI violating effects in the scalar meson
production~\cite{Close2}. That is, the doubly OZI suppressed
process (i.e. doubly disconnected diagram) is assumed to dominate
over the singly OZI suppressed (singly disconnected) process
\cite{Close2}. In contrast, a larger $\Gamma(J/\psi\to \omega
f_0(1710))$ rate over that of $\Gamma(J/\psi\to \phi f_0(1710))$
is naturally accommodated in our scheme without resorting to large
OZI violating effects.

The radiative decay $J/\psi\to \gamma f_0$ is an ideal place to
test the scalar glueball content of $f_0$ since the leading
short-distance mechanism for inclusive $J/\psi\to\gamma+X$ is
$J/\psi\to \gamma+gg$. Its flavor-independence in $J/\psi$ decays
as well as in hadronic and $\gamma\gamma$ productions has been
explored~\cite{lli89}. If $f_0(1710)$ is composed mainly of the
scalar glueball, it should be the most prominent scalar produced
in radiative $J/\psi$ decay. Hence, it is expected that
 \be   \label{1710_1500}
 \Gamma(J/\psi\to \gamma f_0(1710))\gg \Gamma(J/\psi\to \gamma
 f_0(1500)).
 \en
As for $J/\psi\to \gamma f_0(1370)$, it has a destructive
interference between the glueball and $q\bar q$ components. From
the  Particle Data Group \cite{PDG}, $\B(J/\psi\to\gamma
f_0(1710)\to \gamma K\bar K)=(8.5^{+1.2}_{-0.9})\times 10^{-4}$.
Combining with the WA102 measurements \cite{Barberis}: $ \B(f_0\to
\pi\pi)/\Gamma(f_0\to K\bar K)=0.20\pm0.04$ and
$\Gamma(f_0\to\eta\eta)/\Gamma(f_0\to K\bar K)=0.48\pm0.15$ yields
$\B(J/\psi\to\gamma f_0(1710))\sim 1.4\times 10^{-3}$. For
$f_0(1500)$, the BES result $\B(J/\psi\to \gamma
f_0(1500)\to\gamma\pi\pi)=(6.7\pm2.8)\times 10^{-5}$
\cite{BESgammapipi} together with
$\B(f_0(1500)\to\pi\pi)=0.349\pm0.023$ \cite{PDG} gives
$\B(J/\psi\to\gamma f_0(1500))=(2.9\pm1.2)\times 10^{-4}$.
Therefore, $\Gamma(J/\psi\to \gamma f_0(1710))\sim 5\,
\Gamma(J/\psi\to \gamma f_0(1500))$. This is consistent with the
expectation from Eq. (\ref{1710_1500}).

Finally we comment on the strange quark content of $f_0(1370)$.
Although $\rho\rho$ and $4\pi$ are the dominant decay modes of
$f_0(1370)$ \cite{PDG}, it does not necessarily imply that
$f_0(1370)$ is mostly $n\bar n$. In principle, the $s\bar s$
content relative to $n\bar n$ can be determined from the ratio
$R=\Gamma(f_0(1370)\to K\bar K)/\Gamma(f_0(1370)\to \pi\pi)$. If
$f_0(1370)$ is a pure $n\bar n$, $R$ turns out to be 0.23\,. As
noticed in passing, the measured ratio ranges from $1.33\pm0.67$
down to $0.08\pm0.08$. In our scheme, the $s\bar s$ and $u\bar u$
or $d\bar d$ components are similar [see Eq. (\ref{eq:wf})]. Due
to the opposite sign between the glueball and quark contents in
the wavefunction of $f_0(1370)$, $R$ is predicted to be around
0.79 in our scheme.
%
Another ideal place for determining the strange quark component in
$f_0(1370)$ is the decay $D_s^+\to f_0(1370)\pi^+$
\cite{Chengf1370}. If $f_0(1370)$ is purely a $n\bar n$ state, it
can proceed only via the $W$-annihilation diagram. In contrast, if
$f_0(1370)$ has an $s\bar s$ content, the decay $D_s^+\to
f_0(1370)\pi^+$ will receive an external $W$-emission
contribution. In practice, one can compare $\Gamma(D_s^+\to
f_0(1370)\pi^+\to \pi^+\pi^+\pi^-)$ with $\Gamma(D^+\to
f_0(1370)\pi^+\to \pi^+\pi^+\pi^-)$ without the information of
$\B(f_0(1370)\to\pi\pi)$. Unfortunately, the experimental
measurement of $D^+\to f_0(1370)\pi^+\to \pi^+\pi^+\pi^-$ is not
yet available.

\section{Conclusions}
 We have studied the isosinglet scalar mesons
$f_0(1710)$, $f_0(1500)$, $f_0(1370)$ and their mixing. We employ
two recent lattice results as the input for the mass matrix which
is essentially the starting point for the mixing model between
scalar mesons and the glueball; one is the isovector scalar meson
$a_0(1450)$ which displays an unusual property of being nearly
independent of quark mass for quark masses smaller than that of
the strange, and the other is the scalar glueball mass at 1710 MeV
in the quenched approximation. The former implies that, to first
order approximation, flavor SU(3) is a good symmetry for the
scalar mesons above 1 GeV. The latter indicates that the scalar
glueball mass before mixing should be close to 1700 MeV rather
than 1500 MeV.

Our main results are the following: (i) In the SU(3) symmetry
limit, $f_0(1500)$ turns out to be a pure SU(3) octet and is
degenerate with $a_0(1450)$, while $f_0(1370)$ is mainly an SU(3)
singlet with a small mixing with $f_0(1710)$ which is composed
primarily of a scalar glueball. These features remain essentially
unchanged even when SU(3) breaking is taken into account when the
glueball-$q\bar{q}$ mixing is about the same as that between
$q\bar{q}$, i.e. $|y| \sim |x|$. (ii) Sources of SU(3) breaking in
the mass matrix and in the decay amplitudes are discussed. Their
effects are weak and can be treated perturbatively. (iii) Chiral
suppression in the scalar glueball decay into two pseudoscalar
mesons is essential for explaining the width and strong decays of
$f_0(1710)$. (iv) The observed enhancement of $J/\psi\to \omega
f_0(1710)$ production relative to $\phi f_0(1710)$ in hadronic
$J/\psi$ decays and the copious $f_0(1710)$ production in
radiative $J/\psi$ decays lend further support to the prominent
glueball nature of $f_0(1710)$.

\vskip 2.0cm \acknowledgments  We wish to thank M. Chanowitz,
Xiao-Gang He, Shan Jin, J. Schechter and Q. Zhao for useful
discussions. This research was supported in part by the National
Science Council of R.O.C. under Grant Nos. NSC94-2112-M-001-016,
NSC94-2112-M-001-023, NSC94-2811-M-001-059, and the USDOE grant
DE-FG05-84ER40154.

\appendix
\section{Mixing matrix of neutral scalar mesons}
In this appendix we collect the mixing matrices of the isosinglet
scalar mesons $f_0(1710)$, $f_0(1500)$ and $f_0(1370)$ that have
been proposed in the literature. A typical result of the mixing
matrices obtained by Amsler, Close and Kirk~\cite{Close1}, Close
and Zhao~\cite{Close2}, and He {\it et al.}~\cite{He} is the
following
 \be \label{eq:Close}
 \left(\matrix{ f_0(1370) \cr f_0(1500) \cr f_0(1710) \cr}\right)=
\left( \matrix{ -0.91 & -0.07 & 0.40 \cr
                 -0.41 & 0.35 & -0.84 \cr
                0.09 & 0.93 & 0.36 \cr
                  } \right)\left(\matrix{|N\ra \cr
 |S\ra \cr |G\ra \cr}\right),
 \en
taken from \cite{Close2}.

A common feature of these analyses is that $M_S>M_G>M_N$ with
$M_G$ close to 1500 MeV and $M_S-M_N$ of the order $200\sim 300$
MeV. Furthermore, $f_0(1710)$ is considered mainly as a $s\bar s$
state, while $f_0(1370)$ is dominated by the $n\bar n$ content and
$f_0(1500)$ is composed primarily of a glueball with possible
large mixing with $q\bar q$ states.

Based on the lattice calculations, Lee and Weingarten \cite{Lee}
found that $f_0(1710)$ to be composed mainly of the scalar
glueball, $f_0(1500)$ is dominated by the $s\bar s$ quark content,
and $f_0(1370)$ is mainly governed by the $n\bar n$ component, but
it also has a glueball content of 25\%. Their result is
 \be \label{eq:Lee}
 \left(\matrix{ f_0(1370) \cr f_0(1500) \cr f_0(1710) \cr}\right)=
\left( \matrix{ 0.819(89) & 0.290(91) & -0.495(118) \cr
                 -0.399(113) & 0.908(37) & -0.128(52) \cr
                0.413(87) & 0.302(52) & 0.859(54) \cr
                  } \right)\left(\matrix{|N\ra \cr
 |S\ra \cr |G\ra \cr}\right).
 \en
In this scheme, $M_S=1514\pm11$ MeV, $M_N=1470\pm25$ MeV and
$M_G=1622\pm29$ MeV.

With the chiral Lagrangian approach, Giacosa {\it et al.}
\cite{Giacosa} performed a fit to the experimental masses and
decay widths of $f_0(1710)$, $f_0(1500)$ and $f_0(1370)$ and found
four possible solutions, depending on whether the direct decay of
the glueball component is considered. The first two solutions are
obtained without direct glueball decay:
  \be \label{eq:chiralA}
 \left(\matrix{ f_0(1370) \cr f_0(1500) \cr f_0(1710) \cr}\right)=
\left( \matrix{ 0.86 & 0.24 & 0.45 \cr
                 -0.45 & -0.06 & 0.89 \cr
                -0.24 & 0.97 & -0.06 \cr
                  } \right)\left(\matrix{|N\ra \cr
 |S\ra \cr |G\ra \cr}\right),
 \en
and
  \be \label{eq:chiralb}
 \left(\matrix{ f_0(1370) \cr f_0(1500) \cr f_0(1710) \cr}\right)=
\left( \matrix{ 0.81 & 0.19 & 0.54 \cr
                 -0.49 & 0.72 & 0.49 \cr
                -0.30 & 0.67 & -0.68 \cr
                  } \right)\left(\matrix{|N\ra \cr
 |S\ra \cr |G\ra \cr}\right),
 \en
while the last two solutions are phenomenological fits with direct
glueball decay:
  \be \label{eq:chiralC}
 \left(\matrix{ f_0(1370) \cr f_0(1500) \cr f_0(1710) \cr}\right)=
\left( \matrix{ 0.79 & 0.26 & 0.56 \cr
                 -0.58 & 0.02 & 0.81 \cr
                -0.20 & 0.97 & -0.16 \cr
                  } \right)\left(\matrix{|N\ra \cr
 |S\ra \cr |G\ra \cr}\right),
 \en
 and
   \be \label{eq:chiralD}
 \left(\matrix{ f_0(1370) \cr f_0(1500) \cr f_0(1710) \cr}\right)=
\left( \matrix{ 0.82 & 0.57 & -0.07 \cr
                 -0.57 & 0.82 & \sim 0 \cr
                -0.06 & 0.04 & 0.99 \cr
                  } \right)\left(\matrix{|N\ra \cr
 |S\ra \cr |G\ra \cr}\right).
 \en
Among those four solutions, (\ref{eq:chiralA}) and
(\ref{eq:chiralC}) are similar to the mixing matrix
(\ref{eq:Close}).

A solution in which $f_0(1710)$ is dominated by a glueball state
is also found by Burakovsky and Page \cite{Burakovsky}
 \be \label{eq:BP}
 \left(\matrix{ f_0(1370) \cr f_0(1500) \cr f_0(1710) \cr}\right)=
\left( \matrix{ 0.908(50) & 0.133(50) & -0.397(80) \cr
                 -0.305(80) & 0.860(20) & -0.410(40) \cr
                0.287(50) & 0.493(20) & 0.821(20) \cr
                  } \right)\left(\matrix{|N\ra \cr
 |S\ra \cr |G\ra \cr}\right).
 \en
Although this solution is similar to (\ref{eq:Lee}),
(\ref{eq:chiralD}) and ours in (\ref{eq:wf}), the mass difference
of their $M_S$ and $M_N$ is of order 250 MeV. Consequently, the
mass of the $f_0(1370)$ state is predicted to be 1218 MeV in
\cite{Burakovsky}. In our case, $M_S$ is larger than $M_N$ by only
$\sim 25$ MeV which reflects the result from the recent lattice
calculation~\cite{Mathur}.


\end{document}